\documentclass[prl,reprint,twocolumn,showpacs,preprintnumbers,amsmath,amssymb,amsfonts,superscriptaddress]{revtex4}

\usepackage{graphicx}
\usepackage{dcolumn}
\usepackage{bm}
\usepackage{color}
\usepackage{amsmath,amsthm,amssymb,bm}
\usepackage{mathrsfs}
\usepackage{comment}
\usepackage{empheq}

\newcommand{\Rb}{^{\text{87}}\text{Rb}}

\newcommand{\R}{\mathbf{r}}

\newcommand{\integral}{\int\text{d}^3\mathbf{r}}

\begin{document}


\title{Geometrically Frustrated Coarsening Dynamics in Spinor Bose-Fermi Mixtures}
\author{Nguyen Thanh Phuc}
\affiliation{Center for Emergent Matter Science (CEMS), RIKEN, Wako, Saitama 351-0198, Japan}
\author{Tsutomu Momoi}
\affiliation{Center for Emergent Matter Science (CEMS), RIKEN, Wako, Saitama 351-0198, Japan}
\affiliation{Condensed Matter Theory Laboratory, RIKEN, Wako, Saitama 351-0198, Japan}
\author{Shunsuke Furukawa}
\affiliation{Department of Physics, University of Tokyo, 7-3-1 Hongo, Bunkyo-ku, Tokyo 113-0033, Japan}
\author{Yuki Kawaguchi}
\affiliation{Department of Applied Physics, Nagoya University, Furo-cho, Chikusa-ku, Nagoya 464-8603, Japan}
\author{Takeshi Fukuhara}
\affiliation{Center for Emergent Matter Science (CEMS), RIKEN, Wako, Saitama 351-0198, Japan}
\author{Masahito Ueda}
\affiliation{Department of Physics, University of Tokyo, 7-3-1 Hongo, Bunkyo-ku, Tokyo 113-0033, Japan}
\affiliation{Center for Emergent Matter Science (CEMS), RIKEN, Wako, Saitama 351-0198, Japan}
\date{\today}

\begin{abstract}
Coarsening dynamics theory has successfully described the equilibration of a broad class of systems. By studying the relaxation of a periodic array of microcondensates immersed in a Fermi gas which can mediate long-range spin interactions to simulate frustrated classical magnets, we show that coarsening dynamics can be suppressed by geometrical frustration. The system is found to eventually approach a metastable state which is robust against random field noise and characterized by finite correlation lengths with the emergence of topologically stable $\mathrm{Z}_2$ vortices. We find universal scaling laws with no thermal-equilibrium analog that relate the correlation lengths and the number of vortices to the degree of frustration in the system.
\end{abstract}


\maketitle

\textit{Introduction}.--Coarsening dynamics theory~\cite{Hohenberg77, Bray02} has been developed to describe the phase-ordering kinetics following a quench such as a ferromagnet suddenly quenched below the Curie point, a binary alloy undergoing phase separation, and a spinor Bose gas quenched across a phase transition~\cite{Vengalattore10, Guzman11, Mukerjee07}. An important element of this theory is the hypothesis that domain structures and correlation functions at different times in the equilibration process differ only in the overall length scale, and that this length scale grows in time following the power law $\xi(t)\propto t^{1/z}$ with $z$ being the dynamical critical exponent. Such scaling has been examined numerically and experimentally in several models of relaxation dynamics that differ in the symmetry of the order parameter and in the presence of conserved quantities.

Frustration, on the other hand, has long been among the most interesting and challenging issues in condensed matter physics~\cite{Pauling35, Wannier50, Diep-book, Lacroix-book}. Geometrical frustration arises, for example, in a triangular lattice with an antiferromagnetic interaction, where spins cannot align in any energetically favored antiparallel configuration and must instead compromise between competing interactions~\cite{Wannier50, Moessner06}. Magnetic frustration gives rise to a huge degeneracy in the classical ground-state manifold of the system, leading to exotic phases such as spin ice with a macroscopically large residual entropy at zero temperature~\cite{Harris97, Ramirez99, Bramwell01} and spin liquids, in which constituent spins are highly correlated yet strongly fluctuate down to absolute zero~\cite{Anderson73, Fazekas74, Balents10}. Frustrated spin systems also provide a platform for various emergent phenomena such as hidden spin nematic order~\cite{Andreev84, Chandra91}, extended criticality~\cite{Huse03, Hermele04, Henley05} and magnetic monopoles in spin ice~\cite{Castelnovo08, Morris09, Fennell09}. The presence of geometrical frustration is often diagnosed by its susceptibility fingerprint in thermodynamic measurements~\cite{Ramirez94, Balents10}.

In this Letter, by studying the relaxation of a periodic array of microcondensates immersed in a cloud of fermionic atoms which can mimic frustrated classical magnets, we show that the coarsening dynamics can be suppressed by geometrical frustration. The system then approaches a metastable state which has the same local order as the ground state but with finite correlation lengths. It is remarkable that this frustration-induced metastable state is robust against both a random field noise and a small tunneling of atoms between microcondensates. Unlike conventional ultracold atomic systems where the small superexchange interaction between two neighboring atomic spins is used~\cite{Duan03, Trotzky08}, the spinor Bose-Fermi mixtures here can provide a platform to create long-range spin interactions between microcondensates that can extend beyond nearest-neighbor (NN) sites. The interactions are generated through the fermionic medium and enhanced in strength by Bose condensation. The sign and magnitude of the spin interactions can be tuned by varying the densities of fermions and bosons, allowing for an antiferromagnetic interaction, the needed element for magnetic frustration. Compared with the Ising~\cite{Simon11}, XY~\cite{Struck11}, and anisotropic XXZ~\cite{dePaz13, Yan13} antiferromagnets, all of which have been simulated by ultracold atoms, our system realizes an isotropic Heisenberg antiferromagnetic spin model. By varying the strength of next-nearest-neighbor (NNN) interaction that lifts the macroscopic degeneracy in highly frustrated kagome lattice, the correlation lengths of the metastable state can be changed, allowing us to investigate its universal critical properties. In particular, we find new scaling laws with no thermal equilibrium analog that relate the correlation lengths to the degree of frustration in the system. Furthermore, we find that the metastable state characterized by finite correlation lengths contains $\mathrm{Z}_2$ vortices which are topologically stable in triangular and kagome lattices, both of which have been realized for ultracold atoms~\cite{Becker10, Jo12}. The formation of $\mathrm{Z}_2$ vortices can be directly observed in our system with a spin-resolved measurement~\cite{Higbie05}.

\textit{System}.--Consider a two-dimensional periodic array of microcondensates in an optical lattice immersed in a degenerate Fermi gas in a harmonic trap. We assume that the spatial variation of the harmonic trapping potential is smooth over the length scale of the inverse Fermi wavenumber $k_\mathrm{F}^{-1}$ so that the Fermi gas can be regarded as uniform. For the sake of concreteness, we consider spin-1 $\Rb$ BECs and spin-1/2 $^6$Li fermions~\cite{Silber05, Deh08, Li08}. The interaction between bosons and fermions can be decomposed as $V(\R_1,\R_2)=\delta(\R_1-\R_2)\left[g_0\hat{1}+g_1\hat{\mathbf{F}}_1\cdot\hat{\mathbf{F}}_2\right]$, where the coupling constants $g_0$ and $g_1$ are functions of the scattering lengths and $\hat{\mathbf{F}}$ denotes the spin operator. The spin-exchange interaction governs the spin dynamics of the system. We consider a typical case in which the interaction energies are much smaller than the Fermi energy. By using the Schrieffer-Wolff transformation~\cite{Schrieffer66} to adiabatically eliminate the virtual particle-hole excitations in the Fermi gas, we obtain an effective interaction between microcondensates as $\hat{V}_\mathrm{eff}=-V_0\int\text{d}\R\int\text{d}\R'\lambda(k_\mathrm{F}|\R-\R'|)\hat{\mathbf{F}}(\R)\cdot\hat{\mathbf{F}}(\R')$~\cite{Supp-Info},
where the kernel $\lambda$ is the same as that of the RKKY interaction in magnetic metals~\cite{Ruderman54, Kasuya56, Yosida57}.

As the typical size of a microcondensate is much smaller than the spin healing length, the single-mode approximation is valid~\cite{Chang05, Black07, Liu09}. This implies that the three spin components have the same spatial distribution, and thus a microcondensate at lattice site $j$ is characterized by an order parameter $\boldsymbol{\psi}_j=\sqrt{N_\mathrm{b}}(\chi_{1,j},\chi_{0,j},\chi_{-1,j})^\mathrm{T}$, where $N_\mathrm{b}$ is the total number of particles in a micro-condensate and the spinor order parameter is normalized to unity: $\sum_{m=-1}^1|\chi_{m,j}|^2=1$. If the spatial distribution of particles in a microcondensate is described by a wavefunction $\phi(\R)$ localized around a lattice site, we can express the interaction energy in terms of  the spinor order parameter as $V(\{\chi_{m,j}\})=J_0\sum_j\mathbf{S}_j^2+\sum_{(i,j)}J_{ij}\mathbf{S}_i\cdot\mathbf{S}_j$, where $\mathbf{S}_j=\sum_{m,n=-1}^1 \chi_{m,j}^* \mathbf{f}_{mn}\chi_{n,j}$ with $\mathbf{f}_{mn}$ being the matrix element of the spin-1 matrix vector. The coupling constants $J_0$ and $J_{ij}$ are functions of $\lambda(x)$ and $\phi(\R)$~\cite{Supp-Info}. Each microcondensate becomes a giant spin and the spin interactions are enhanced by the Bose-Einstein condensation. Their signs and magnitudes can be tuned by varying the density $n_\mathrm{f}$ of fermions, the spatial extent $d$ of a microcondensate, and the lattice constant $a$. For example, if we consider a mixture of $\Rb$ and $^6$Li with $N_\mathrm{b}\simeq1000$ and $n_\mathrm{f}\simeq 5\times10^{9}\,\text{cm}^{-3}$ in a triangular or kagome lattice with $a\simeq 4.6\,\mu\text{m}$ and an isotropic harmonic distribution $\phi(\R)=e^{-|\R|^2/(4d^2)}/(2\pi d^2)^{3/4}$ with $d=k_\mathrm{F}^{-1}/2\simeq 1\,\mu\text{m}$, the on-site, NN, and NNN interactions are estimated to be $J_0/\hbar\simeq -300\,\text{Hz}$, $J_1/\hbar\simeq 70\,\text{Hz}$, and $J_2/\hbar\simeq -7\,\text{Hz}$, respectively. Long-range spin interactions beyond $J_2$ are negligibly small. These coupling constants can be made even larger by, for example, elongating the microcondensates in the direction perpendicular to the 2D lattice. Since $J_0<0$ and $J_1>0$, the microcondensates tend to be polarized locally, and interact with one another by an antiferromagnetic NN interaction.  

\begin{figure}[tbp] 
  \centering
  \includegraphics[width=3in,keepaspectratio]{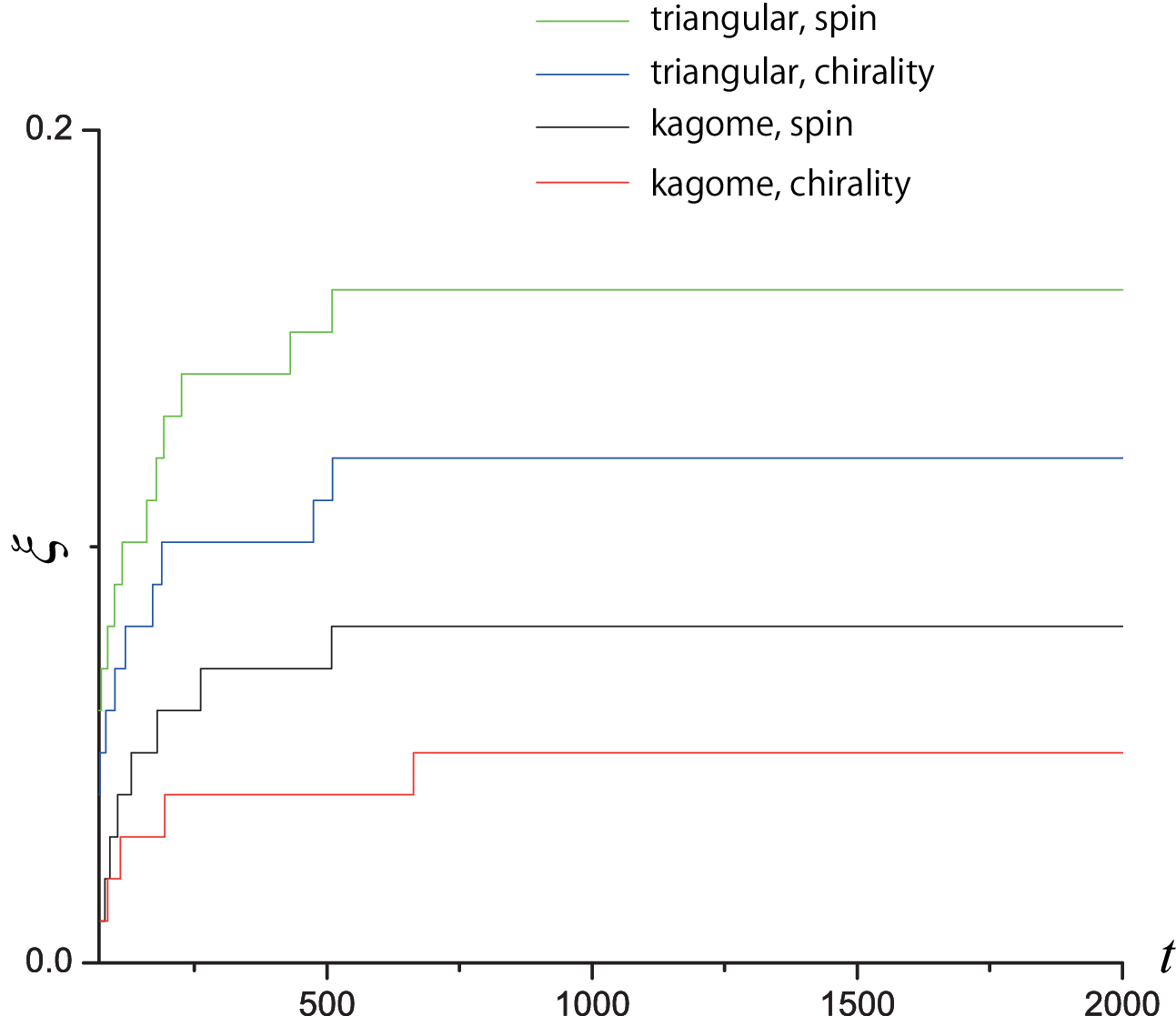}
  \caption{(color online) Time-dependent spin- and chirality-correlation lengths, $\xi_\mathrm{s}$ and $\xi_\mathrm{c}$, for triangular and kagome lattices. Here, time is measured in units of $\hbar/J_1$, where $J_1>0$ is the nearest-neighbor antiferromagnetic interaction, and the correlation lengths are measured in units of the system's size. For the kagome lattice, a next-nearest-neighbor interaction $J_2=-0.1J_1$ is added to lift the degeneracy in the ground-state manifold due to geometrical frustration. The stepwise changes in $\xi_\mathrm{s}$ and $\xi_\mathrm{c}$ are due to the fact that the correlation lengths are only determined by integer multiples of the lattice constant (see the text for details). The early time evolution ($J_1t/\hbar\leq 70$) are not shown because the initial ferromagnetic spin order remains dominant and thus the antiferromagnetic correlation lengths are ill-defined.}
  \label{fig: time dependence of spin and chirality correlation lengths}
\end{figure}

\textit{Frustrated spin dynamics}.--We now study the relaxation dynamics of the spinor microcondensate ensemble. Initially, most of the atoms in the condensates are prepared in the $m_F=1$ Zeeman sublevel, corresponding to the high-energy ferromagnetic state. As the perfect ferromagnetic state is a steady state, there would be no time evolution starting from such a state. To drive the system away from the initial state, we add small fluctuations in the Zeeman sublevel populations which usually arise from experimental realities~\cite{Supp-Info}. Since the atomic interactions are small compared with the critical temperature of the Bose-Einstein condensation and the number of particles are sufficiently large in each microcondensate, the dynamics of the system can be described by the time-dependent Gross-Pitaevskii (GP) equation~\cite{Pethick-book}. In addition to the effective spin interaction above, the coupling between bosons and fermions also leads to a spin relaxation of the microcondensates characterized by a non-local Gilbert damping term $\gamma(|\R-\R'|)$ in the Landau-Lifshitz-Gilbert equation generalized to a spatially inhomogeneous spin system. Similar to the kernel $\lambda$ in the RKKY interaction, the fermion-induced non-local Gilbert damping is an oscillating and rapidly decaying function of distance~\cite{Umetsu12}. Therefore, the dominant contribution to the spin relaxation of a microcondensate comes from the dynamics of its particles, leading to an effective Gilbert damping of $\Gamma=N_\mathrm{b}\integral\, \gamma(\R)|\phi(\R)|^2\sim N_\mathrm{b}g_1^2M_\mathrm{f}^2k_\mathrm{F}^2/\hbar^4$. Using the parameters of the $\Rb$-$^6$Li mixture, we find $\Gamma\sim 0.1$. On the other hand, the spin relaxation of a ferromagnetic BEC can equivalently be taken into account by adding the Gilbert damping coefficient $\Gamma$ to the left-hand side of the GP equation~\cite{Kudo11}, yielding
\begin{align}
(i-\Gamma)\hbar \frac{\text{d}\chi_{m,j}}{\text{d}t}=&\left(2J_0\mathbf{S}_j+\sum_{i\not=j}J_{ij}\mathbf{S}_i\right)\cdot\left(\sum_n\mathbf{f}_{mn}\chi_{n,j}\right).
\label{eq: GP equation with dissipation}
\end{align}
We numerically solve Eq.~\eqref{eq: GP equation with dissipation} to find the spin relaxation dynamics of the system. Here, we use the open boundary condition to simulate realistic experiments, the number of sites in one direction of the Bravais lattice is $L=100$, and the normalization of the order parameter is performed at each discrete time step. Figure~\ref{fig: time dependence of spin and chirality correlation lengths} shows representative time evolutions of spin-correlation length $\xi_\mathrm{s}$ and chirality-correlation length $\xi_\mathrm{c}$ for triangular and kagome lattices, where the chirality vector is defined by $\mathbf{C}_i\equiv(2/3\sqrt{3})(\mathbf{S}_{i,1}\times \mathbf{S}_{i,2}+\mathbf{S}_{i,2}\times\mathbf{S}_{i,3}+\mathbf{S}_{i,3}\times\mathbf{S}_{i,1})$ with $\mathbf{S}_{i,1}, \mathbf{S}_{i,2}, \mathbf{S}_{i,3}$ being magnetizations at three vertices of plaquette $i$ in the anti-clockwise direction. Since the system is in a nonequilibrium state, the correlation functions during the dynamics generically do not follow rigorous exponential functions. The correlation lengths, however, can be evaluated by the distance at which their magnitudes drop to half of the maximum value. It is clear from Fig.~\ref{fig: time dependence of spin and chirality correlation lengths} that spin systems in lattices with geometrical frustration approach a metastable state with finite $\xi_\mathrm{s}$ and $\xi_\mathrm{c}$, in contrast to the standard picture of coarsening dynamics where the correlation lengths grow indefinitely. Remarkably, unlike other metastable states in many-body systems which usually decay to the ground state as a small random field is added, these frustration-induced metastable states turn out to be robust against such a noise. Introducing a small tunneling of atoms between lattice sites does not destroy that metastable state either~\cite{Supp-Info}.

\begin{figure}[tbp] 
  \centering
  \includegraphics[width=3in,keepaspectratio]{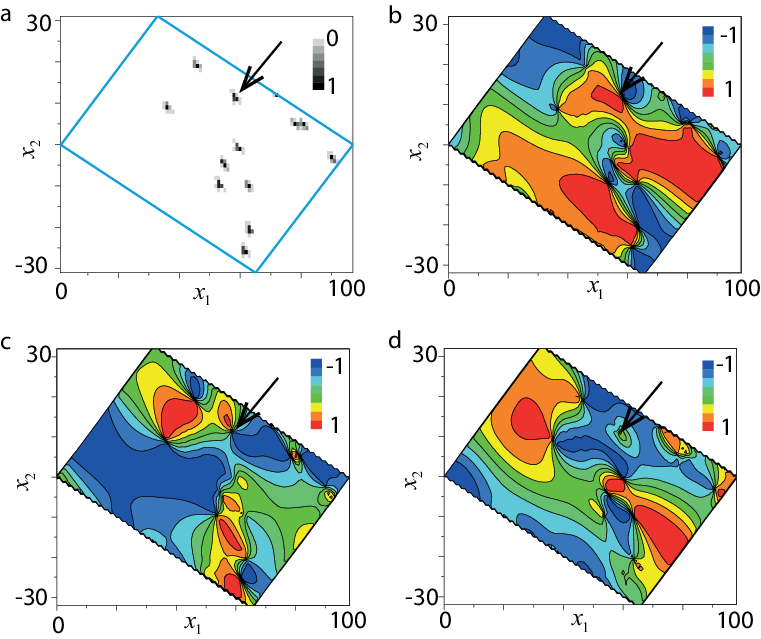}
  \caption{Spatial distributions of (a) winding numbers of $\mathrm{Z}_2$ vortices and three magnetization components (b) $S_x$, (c) $S_y$, (d) $S_z$ in the system in the triangular lattice at time $J_1t/\hbar=100$. The winding number and the magnetization are represented by the gray scale and the color gauge, respectively. Here $x_1$ and $x_2$ are spatial coordinates of the Bravais lattice and the spatial distributions are coarse-grained to the scale of the sublattice where the spin configuration in the ground state becomes homogeneous~\cite{Supp-Info}. The arrows indicate the location of a representative SO(3) vortex.}
  \label{fig: spatial distribution of winding number for triangular lattice}
\end{figure}

It is also evident from Fig.~\ref{fig: time dependence of spin and chirality correlation lengths} that the growth of correlations in the system slows down with increasing geometrical frustration from triangular to kagome lattices. Moreover, the growth of $\xi_\mathrm{c}$ is always slower than that of $\xi_\mathrm{s}$. This can be understood as a collective effect since the formation of a spin chirality of a triangular plaquette involves the magnetizations at three vertices. The finite correlation lengths at long time suggest that spin domains appear in the metastable state. Similar to the domain formation in quench dynamics through a second-order phase transition with spontaneous symmetry breaking~\cite{Sadler06, Kibble76, Zurek85}, the emergence of spin domains found here is expected to accompany topological defects. In particular, as a local order is formed in antiferromagnets in triangular and kagome lattices, the three magnetizations in each plaquette tend to form an angle of 120$^\circ$ with one another~\cite{Supp-Info}. The three magnetizations and the chirality vector of a plaquette then form a tetrahedron, whose free rotation in space yields the SO(3) order-parameter manifold of the system. As the first homotopy group of this manifold is given by $\pi_1(\mathrm{SO(3)})=\mathrm{Z}_2$, there can exist a stable topological defect called $\mathrm{Z}_2$ vortex~\cite{Kawamura84}. To get the spatial distribution of $\mathrm{Z}_2$ vortices in the system, we calculate the winding number made by the spin configuration along a loop connecting three plaquettes, whose spin configurations are identical in the ground state. We use the SU(2) representation of SO(3) rotations, with which a $2\pi$ rotation can be distinguished from no rotation~\cite{Supp-Info}. A representative spatial distribution of the winding numbers of $\mathrm{Z}_2$ vortices for the triangular lattice is shown in Fig~\ref{fig: spatial distribution of winding number for triangular lattice}a, where the generation of $\mathrm{Z}_2$ vortices in the nonequilibrium dynamics is clearly seen. The $\mathrm{Z}_2$ vortices can be directly observed by using a spin-resolved measurement~\cite{Higbie05} of the distribution of three components of the magnetization (Fig~\ref{fig: spatial distribution of winding number for triangular lattice}b--d).

\begin{figure}[tbp] 
  \centering
  \includegraphics[width=3in,keepaspectratio]{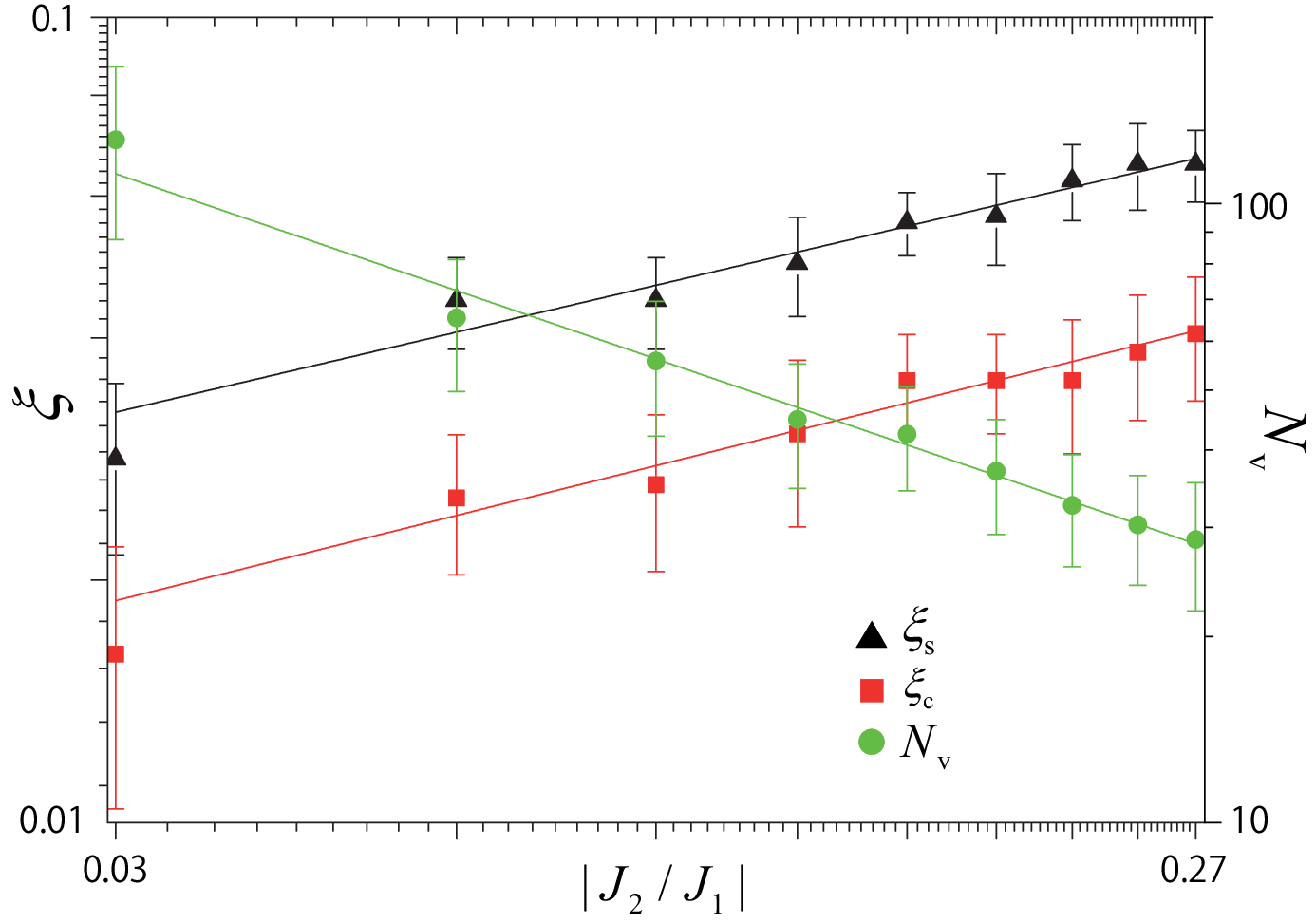}
  \caption{Dependences of the spin-correlation length $\xi_\mathrm{s}$ (black triangles), the chirality-correlation length $\xi_\mathrm{c}$ (red squares) and the number of vortices $N_\mathrm{v}$ (green circles) on the ratio of the next-nearest-neighbor interaction $J_2<0$ to the nearest-neighbor one $J_1>0$ for the system in the kagome lattice. These quantities are evaluated at a fixed evolution time and displayed in the logarithmic scales. The correlation lengths are measured in units of the system size. The averages are taken over ten initial states with random phases. Straight lines show the least-square fittings of the corresponding numerical data.}
  \label{fig: J2 dependence of correlation lengths and number of Z2 vortices}
\end{figure}

While the ground state is uniquely determined for antiferromagnets in the triangular lattice with a NN interaction $J_1>0$, there is a macroscopically large degeneracy in the classical ground-state manifold of the system in the kagome lattice due to geometrical frustration. To lift this degeneracy and to induce a long-range spin order, a NNN interaction $J_2\not=0$ is needed. The $J_2$-dependences of the long-time correlation lengths $\xi_\mathrm{s}$, $\xi_\mathrm{c}$, and the number of vortices $N_\mathrm{v}$ are shown in Fig.~\ref{fig: J2 dependence of correlation lengths and number of Z2 vortices} for $J_2<0$. Here the data are averaged over the random phases of spinor components of the initial state. The error bars involve both the limited precision in determining the correlation lengths due to the discrete lattice structures and the statistical standard devitation due to random initial phases. It is evident that the correlation lengths increase with increasing magnitude of $J_2$ by which frustration is reduced, while the number of vortices $N_\mathrm{v}$ decreases as the spin domains get bigger. A linear relation in logarithmic scales in Fig.~\ref{fig: J2 dependence of correlation lengths and number of Z2 vortices} implies the scaling laws of $\xi_\mathrm{s}$, $\xi_\mathrm{c}$, and $N_\mathrm{v}$ with respect to $|J_2|$. Using the least-square fitting procedure, we find $\xi_\mathrm{s}\sim |J_2|^\alpha$, $\xi_\mathrm{c}\sim |J_2|^\beta$, and $N_\mathrm{v}\sim |J_2|^{-\gamma}$ with $\alpha= 0.33\pm 0.03$, $\beta=0.35\pm 0.04$, and $\gamma=0.63\pm0.03$. The relation of $\alpha\simeq \beta\simeq \gamma/2$ to within their error bars can be understood by the fact that the number of vortices is approximately equal to the area of the system divided by the area of a spin domain which is approximately given by the correlation length squared. 

By varying the Gilbert damping coefficient $\Gamma$ and changing the initial state between ferromagnetic and polar phases, we find that the critical exponents $\alpha$, $\beta$, and $\gamma$ of the metastable state depend on neither $\Gamma$ nor the initial condition to within their error bars~\cite{Supp-Info}. This implies their universality. To make a comparison, it is noteworthy that the ground state of the system always has an infinite correlation length except for a single point of $J_2=0$ where the correlation length vanishes. The  scaling laws and critical exponents $\alpha$, $\beta$ and $\gamma$ are also investigated for the case of $J_2>0$, where the ground state of an antiferromagnet in the kagome lattice  changes from the $\sqrt{3}\times \sqrt{3}$ Neel state to the $\mathbf{q}=0$ Neel state~\cite{Harris92}. It is found that the values of those critical exponents increase as $J_2$ changes its sign from negative to positive. The smaller correlation lengths for $J_2<0$ can be understood qualitatively by looking at the energy landscape of the system as a function of $J_2$~\cite{Supp-Info}. Since the density of states near the ground state for $J_2<0$ turns out to be larger than that for $J_2>0$ with the same modulus, the manifold of metastable states with finite correlation lengths for $J_2<0$ has larger degeneracy and stronger frustration. It is this frustration that suppresses the growth of correlation lengths.

\textit{Conclusion}.--By studying the relaxation dynamics of a periodic array of microcondensates immersed in a Fermi gas, we have shown that the coarsening dynamics can be suppressed by geometrical frustration. Instead of decaying to the ground state, the system is found to approach a metastable state which has the same local order as the ground state but with a finite correlation length. The fluctuation-induced metastable state turns out to be remarkably robust against both random field noise and tunneling of atoms between lattice sites. This metastable state also contains $\mathrm{Z}_2$ vortices, which are topologically stable in triangular and kagome lattices and can be directly observed by spin-resolved measurements. By varying the next-nearest-neighbor spin interaction in kagome lattice, we are able to investigate universal critical properties of the metastable state. In particular, we find new scaling laws relating the metastable state's correlation lengths and the number of vortices to the degree of frustration of the system, which have no thermal equilibrium analog. Although here we consider a system of ultracold atoms, the obtained results are universal at least qualitatively and can be applied to any other frustrated classical spin. Furthermore, by using the same setup with fermionic atoms in larger hyperfine-spin states, the dynamics of a system with exotic spin interactions that do not exist in conventional condensed matters can be explored. 

\begin{acknowledgments}
This work was supported by KAKENHI Grant Nos. JP23540397, JP25800225, JP26287088, JP15K17726, and JP16K05425 from the Japan Society for the Promotion of Science, a Grant-in-Aid for Scientific Research on Innovative Areas "Topological Materials Science" (KAKENHI Grant Nos. JP15H05855 and JP16H00989), the Photon Frontier Network Program from MEXT of Japan, the Mitsubishi Foundation, and the ImPACT Program of Council for Science, Technology and Innovation (Cabinet Office, Government of Japan).
\end{acknowledgments}


\end{document}